\documentclass[twocolumn]{IEEEtran} 
\usepackage{amssymb,amsmath}
\usepackage{graphicx,float}
\DeclareMathOperator{\Tr}{Tr}
\newcounter{append}
\def\appendix{\par
	  \setcounter{section}{0}\setcounter{subsection}{0}
	   \setcounter{equation}{0}
		\def\thesection{\Alph{section}}\section*{Appendix \Alph{append}}
		 \def\theequation{\thesection\Alph{append}.\arabic{equation}}
}
\def\BibTeX{{\rm B\kern-.05em{\sc i\kern-.025em b}\kern-.08em
	T\kern-.1667em\lower.7ex\hbox{E}\kern-.125emX}}

\begin{document}

\title{Robust Source Localization in a Random Shallow Water Channel}

\author{Alexander Sazontov, Ivan Smirnov and Alexander Matveyev
\thanks{A.~Sazontov, I. Smirnov and A.~Matveyev
are with the 
Institute of Applied Physics,
Russian Academy of Science, Nizhny Novgorod,
Russia.}}



\maketitle

\begin{abstract}
This paper addresses
source localization problem in a random shallow water channel.
We present an extension of the generalized MUSIC method to the case, 
when the signal correlation matrix is imprecisely known.
The algorithm is validated by 
its application to the experimental data observed in the Barents Sea.
It has been found that the approach proposed demonstrates its excellent performance.
\end{abstract}


\section{Introduction}

\PARstart{T}{he} source localization in shallow water by means of matched field processing has been an area of active interest.
However, this approach is known to be extremely sensitive to errors in the assumed environmental conditions
and has poor robustness against the underwater channel uncertainty which limits
its application in practical engineering.

At relatively short propagation distances when the spatial signal coherence length is large compared to the array aperture,
the deterministic environmental
mismatch (which arises  due to imprecise knowledge of sound speed profile, water depth, and bottom characteristics) is the primary cause of failure.
To minimize the effects of the uncertainties in waveguide parameters,
several adaptive methods (described, for example, in Refs.~\cite{VT-IV,Li})
have been proposed to improve the performance of the source localization. 

With increasing transmission path it is also necessary to take into account the loss of coherence
which results from multiple sound scattering by random inhomogeneities of the underwater channel.
The appearance of the amplitude and phase fluctuations of acoustic wavefield
can be interpreted as a sort of multiplicative noise.
In the presence of such ``noises'', a rank-one signal model assumption is not applicable
and conventional  algorithms fail to give consistent matching with real data.
In such a situation,  matching can be performed only in a statistical sense.
\nocite{LaLa,PauKai88,MorS,Gershman97,GMSV99,Gershman00,BeViStoGe00}
As a result, the knowledge of the spatial signal correlation matrix along the array aperture
is of the uppermost importance
to optimize the signal processing techniques and, therefore,
to decrease a coherence-induced degradation of the array processor performance
(see, e.g.,\cite{LaLa}--\cite{BeViStoGe00}).
However, the complete information about a stochastic underwater channel
is rarely available under most practical scenarios
and effective estimation algorithms must be developed to improve robustness against statistical mismatch.

The basic concept of the robust adaptive array processing for general-rank signal models
address the situation when the desired signal covariance matrix is not known precisely. It
assumes the norm bounded mismatch 
and uses the worst-case principle (see, e.g., the review article~\cite{Vor13}  and the references therein).
In this context, the development of the corresponding methods for
solving the source localization problem in a random shallow water is of special interest.

In this paper, a robust version of a subspace-based estimator for source range and depth
in scattering shallow water environments is constructed.
The real data processing results observed in the Barents Sea demonstrate essential  performance improvements that can be achieved by means of the approach proposed.

The body of this paper is organized as follows.  
In Section~II we introduce a general-rank signal model and formulate the source localization
algorithm  in the case of imperfect spatial coherence.
Next, in Section III, we develop an extended version
of the generalized MUSIC method, that is robust to a mismatched signal correlation matrix,
and derive a  closed-form solution  to  this problem.
The resulting algorithm is then tested  in Section IV  on actual data collected in the Barents sea.
Finally, a few concluding remarks are given in Section V.

\section{Background}

Let a  point source be located at depth $z_0$ and range $r_0$ and emit a narrow-band signal $s(t)$,
where $s(t)$ is considered to be a stationary, zero-mean random process.
This signal passing through a random  channel is registered by a
vertical receiving array of $N$ sensors located at the depths $\{z_j\}_{j=1}^N$.

Assuming narrow-band processing, the $N\times 1$ observation vector $\boldsymbol{x}(t_l)$
at time instant $t_l$  can be written as
$$
\boldsymbol{x}(t_l)=s(t_l)\mathbf{e}(\pmb{\theta})+{\boldsymbol{n}}(t_l).
$$
Here, $L$ denotes the number of data snapshots available,
$\mathbf{e}(\pmb{\theta})$ is the signal vector depending on unknown source position
$\pmb{\theta}=(r_0,z_0)^{T}$ of the form
$$
\mathbf{e}(r_0,z_0)=\Bigl[G(0,z_1|\boldsymbol{r}_0,z_0),
\cdots,
G(0,z_N|\boldsymbol{r}_0,z_0)\Bigr]^{T},
$$
where $G(0,z_j|\boldsymbol{r}_0,z_0)$
is the Green's function of the Helmholtz equation, 
and $\boldsymbol{n}$ is a vector of additive noise.  
(The superscript  $T$ stands for transpose.)

The signal and noise are assumed to be independent of each other and 
the covariance matrix of the array output can be represented as
\begin{equation*}
\boldsymbol{\Gamma}_{\boldsymbol{x}}\!=\!\sigma_s^2\mathbf{R}_{\mathbf{s}}(%
\pmb{\theta})+\boldsymbol{\Gamma}_{\boldsymbol{n}},
\mathbf{R}_{\mathbf{s}}(\pmb{\theta})\!=\!\bigl.\left<\mathbf{e}(\pmb{\theta})
\mathbf{e}^{+}(\pmb{\theta})\right>\bigr/\overline{I}_s,
\boldsymbol{\Gamma}_{\boldsymbol{n}}=\!\left<\boldsymbol{n}\boldsymbol{n}^{+}\right>,
\end{equation*}
where  $\sigma_s^2=\left<|s(t)|^2\right>\!\overline{I}_s$ denotes the input signal power
$\overline{I}_s=\mathrm{Tr}\,[\left<{\bf e}{\bf e}^{+}\right>]$
is the mean intensity of the sound field on the array aperture, and
$\mathbf{R}_{\bf s}(\pmb{\theta})$ is the $N\times N$ signal correlation matrix
(whose rank can be between $1$ and $N$) %
which satisfies the norm constraint $\mathrm{Tr}\,({\bf R}_{\bf s})=1$.
(The superscript $^{+}$ denotes the conjugate transpose,
the angular brackets $<\cdots>$ indicate  ensemble averaging,
and $\mathrm{Tr}(\cdot)$ stands for the trace operator.)

In practical situations, the exact data covariance matrix is unavailable
and is replaced by its sample estimate
\begin{equation*}
\hat{\boldsymbol{\Gamma}}_{\boldsymbol{x}}=\dfrac1L\,\sum\limits_{l=1}^{L}
\boldsymbol{x}(t_l)\boldsymbol{x}^{+}(t_l),
\end{equation*}
where $L$ is the number of data vectors in the observation period.
The problem of interest is to estimate the source position 
from the data matrix $\hat{\boldsymbol{\Gamma}}_{\boldsymbol{x}}$.

Most approaches to robust adaptive beamforming
are based on the eigenvalue decomposition of  
$\hat{\boldsymbol{\Gamma}}_{\boldsymbol{x}}$:
\begin{equation*}
\hat{\boldsymbol{\Gamma}}_{\boldsymbol{x}}=
\hat{\pmb{\Psi}}_{\bf s}\hat{\pmb{\Lambda}}_{\bf s}
\hat{\pmb{\Psi}}_{\bf s}^{+}+\hat{\pmb{\Psi}}_{\boldsymbol{n}}
\hat{\pmb{\Lambda}}_{\boldsymbol{n}}\hat{\pmb{\Psi}}_{\boldsymbol{n}}^{+},\ 
\hat{\lambda}_1\geqslant
\cdots\geqslant\hat{\lambda}_J\geqslant
\cdots\geqslant\hat{\lambda}_N,
\end{equation*}
where the $N\times J$ matrix $\hat{\pmb{\Psi}}_{\bf s}$ and $N\times (N-J)$ matrix $\hat{\pmb{\Psi}}_{\boldsymbol{n}}$
contain the signal subspace eigenvectors
of $\hat{\boldsymbol{\Gamma}}_{\boldsymbol{x}}$ and the noise subspace eigenvectors, respectively,
while  the diagonal matrices $\hat{\pmb{\Lambda}}_{\bf s}$ and $\hat{\pmb{\Lambda}}_{\boldsymbol{n}}$ contain the $J$
largest and $N-J$ smallest eigenvalues, respectively.
This representation is extensively used in the description and implementation of the subspace-based estimation algorithms.

One of the most popular and most powerful 
superresolution methods is the 
MUSIC 
exploiting the fact that the actual steering vector is orthogonal to noise subspace.
In the case when the received signal is perfectly coherent along the array aperture
the output power of the traditional MUSIC processor is defined as~\cite{Schimdt86}:
\begin{equation}
\label{MUSIC}
P_{\text{MUSIC}}(\pmb{\theta})=\dfrac{{\bf e}^{+}(\pmb{\theta})\,{\bf e}(\pmb{\theta})}%
{{\bf e}^{+}(\pmb{\theta})\hat{\pmb{\Pi}}_{\boldsymbol{n}}
{\bf e}(\pmb{\theta})},
\end{equation}
where $\hat{\pmb{\Pi}}_{\boldsymbol{n}}=\hat{\pmb{\Psi}}_{\boldsymbol{n}}
\hat{\pmb{\Psi}}_{\boldsymbol{n}}^{+}$
is the estimated projector onto the noise subspace.
The  corresponding technique can be viewed as
an un-weighted noise subspace fitting method~\cite{VibOt91} where the source position is found as the highest peak of $P_{\text{MUSIC}}(\pmb{\theta})$.


To generalize MUSIC in the situation of  imperfect spatial coherence several closely related
subspace-based methods have been proposed for the problem of interest.
Among others, the so-called  DSPE~\cite{VaChaKa95} and  DISPARE~\cite{MeStoWo96}
algorithms are widely applicable in the signal processing literature.
It is worth noting that these methods rely on restrictive hypothesis that
most of signal energy is concentrated in a few eigenvalues of the array covariance matrix.
Another related approach is presented in Ref.~\cite{Beng01}, where
a more general class of weighted subspace (but very high-complexity) algorithms
for consistent estimation of source parameters from a possibly full rank data model is suggested.

However, all of these techniques are based on a priori knowledge of the signal correlation matrix characterizing the loss of coherence along the array aperture.
In practice, this assumption may be unrealistic: as mentioned above,in the presence of random uncertainties,
there is always a certain mismatch between the actual and presumed values of the signal matrix,
which results in a decrease of the localization performance.

The algorithm under consideration in this paper
has been originally developed in~\cite{VaChaKa95} and leads to the following  DSPE criterion
\begin{equation}
\label{MUSIC:general}
\hat{\pmb{\theta}}=
\arg\max\limits_{\pmb{\theta}}P_{\text{GMUSIC}}(\pmb{\theta}),\
P_{\text{GMUSIC}}(\pmb{\theta})=
\left[{{\rm Tr}\bigl\{\hat{\pmb{\Pi}}_{\boldsymbol{n}}
{\bf R}_{\bf s}({\pmb{\theta}})\bigr\}}\right]^{-1}.
\end{equation}
For the studied scenarios,
it exploits the approximate orthogonality between the estimated pseudo-noise subspace,
from the sample covariance matrix, and the theoretical pseudo-signal subspace.
In particular, in the the rank-one case, this estimator reduces to the ordinary MUSIC method~\eqref{MUSIC}.

Below, we present an extended version of~\eqref{MUSIC:general} that is robust to mismatched signal correlation matrix.

\section{The generalized robust MUSIC algorithm} 

To provide robustness against  statistical mismatch
let us represent the positive definite Hermitian signal matrix
${\mathbf{R}}_{\mathbf{s}}({\pmb{\theta}})$ as
${\mathbf{R}}_{\mathbf{s}}=\mathbf{D}_0\mathbf{D}_0^{+}$.
Note that the matrix $\mathbf{D}_0$ (the square root of ${\mathbf{R}}_{\mathbf{s}}$)
satisfies the normalization condition:
$\mathrm{Tr}(\mathbf{D}_0\mathbf{D}_0^{+})=\mathrm{Tr}\,{\mathbf{R}}_{\mathbf{s}}=1$.

Then, we assume that the  actual matrix $\mathbf{D}$ differs from its presumed value  $\mathbf{D}_0$
by some unknown covariance matrix error and the corresponding matrix mismatch is bounded
by a given constant $\varepsilon$: $\lVert\mathbf{D}-\mathbf{D}_0\rVert^2_F\leqslant\!\varepsilon$,
where $\lVert\cdot\rVert_F$ indicates the Frobenius norm.
With such constraints, the optimum robust matrix $\hat{\mathbf{D}}$ can be
estimated by maximizing the output power~\eqref{MUSIC:general} 
(or equivalently by minimizing the denominator of~\eqref{MUSIC:general}):
\begin{equation*}
\min\limits_{\mathbf{D}} \bigl\{\mathrm{Tr}[\mathbf{D}^{+}\hat{\boldsymbol{%
\Pi}}_{\boldsymbol{n}}\mathbf{D}]\bigr\}\ \text{ s.t. }\ \lVert%
\mathbf{D}-\mathbf{D}_0\rVert^2_F\leqslant\!\varepsilon, \ \mathrm{Tr}(%
\mathbf{D}\mathbf{D}^{+})=1.
\end{equation*}
This problem can be solved by using the Lagrange multiplier method 
based on the function
\begin{equation*}
L(\mathbf{D},\mu,\nu)=\mathrm{Tr}[\mathbf{D}^{+}
\hat{\boldsymbol{\Pi}}_{\boldsymbol{n}} \mathbf{D}]+\mu\Bigl[\lVert\mathbf{D}-
\mathbf{D}_0\rVert^2_F-\varepsilon\Bigr]+
\nu\Bigl[\lVert\mathbf{D}\rVert^2_F-1\Bigr],
\end{equation*}
where $\mu$  and $\nu$ are the real-valued Lagrange multipliers.
The above function can be rewritten equivalently as
\begin{multline*}
L(\mathbf{D},\mu,\nu)=\mathrm{Tr}\Bigl\{\bigl[\mathbf{D}-
\mu(\hat{\boldsymbol{\Pi}}_{\boldsymbol{n}}+\nu\mathbf{I})^{-1}\mathbf{D}_0\bigr]^{+}\\
\times
(\hat{\boldsymbol{\Pi}}_{\boldsymbol{n}}+ \nu\mathbf{I}) \bigl[\mathbf{D}%
-\mu(\hat{\boldsymbol{\Pi}}_{\boldsymbol{n}}+ \nu\mathbf{I})^{-1}\mathbf{D}_0%
\bigr]\Bigr\}- \\
- \mu^2\mathrm{Tr}\bigl[\mathbf{D}_0^{+}(\hat{\pmb{\Pi}}_{\boldsymbol{n}}+
\nu\mathbf{I})^{-1}\mathbf{D}_0\bigr]+\mu(2-\varepsilon)-\nu.
\end{multline*}
Minimization of $L$ with respect to $\mathbf{D}$ and $\mu$ gives
\begin{equation}
\label{D:rob}
\hat{\mathbf{D}}=\mu(\hat{\boldsymbol{\Pi}}_{\boldsymbol{n}}+
\nu\mathbf{I})^{-1}\mathbf{D}_0,\quad \mu=\dfrac{1-\varepsilon/2} {\mathrm{Tr\bigl[%
\mathbf{D}_0^{+}(\hat{\boldsymbol{\Pi}}_{\boldsymbol{n}}+ \nu\mathbf{I})^{-1}%
\mathbf{D}_0\bigr]}}
\end{equation}
and Lagrange multipliers $\nu$ can be found
by substituting~\eqref{D:rob} into $\lVert{\bf D}\rVert^2_F=1$. That is,  $\nu$  can be obtained
by solving
\begin{equation}  \label{nu:genmus}
\dfrac{\mathrm{Tr\bigl[\mathbf{D}_0^{+}(\hat{\boldsymbol{\Pi}}_{\boldsymbol{n}}+
\nu\mathbf{I})^{-2}\mathbf{D}_0\bigr]}}{\mathrm{Tr^2\bigl[\mathbf{D}%
_0^{+}(\hat{\boldsymbol{\Pi}}_{\boldsymbol{n}}+
\nu\mathbf{I})^{-1}\mathbf{D}_0\bigr]}}=\dfrac1{(1-\varepsilon/2)^2}.
\end{equation}

To derive a closed form solution to $\nu$ we employ
the matrix inverse lemma that
$$
\bigl(\mathbf{A}+\mathbf{B}\mathbf{C}\bigr)^{-1}=\mathbf{A}^{-1}-
\mathbf{A}^{-1}\mathbf{B} \bigl(\mathbf{I}+\mathbf{C}\mathbf{A}^{-1}%
\mathbf{B}\bigr)^{-1}\mathbf{C}\mathbf{A}^{-1}.
$$
By letting
$\mathbf{A}=\nu\mathbf{I}$, $\mathbf{B}=\hat{\pmb{\Psi}}_{\boldsymbol{n}}$,
$\mathbf{C}=\hat{\pmb{\Psi}}_{\boldsymbol{n}}^{+}$
and taking into account that
 $\hat{\pmb{\Psi}}_{\boldsymbol{n}}^{+}\,\hat{\pmb{\Psi}}_{\boldsymbol{n}}=\mathbf{I}$, 
one gets
\begin{equation}  \label{inversion}
\bigr(\hat{\pmb{\Psi}}_{\boldsymbol{n}}\hat{\pmb{\Psi}}_{\boldsymbol{n}}^{+}+
\nu\mathbf{I}\bigr)^{-1}=\dfrac1{\nu}\,\Bigl[\mathbf{I}-
\dfrac{\hat{\pmb{\Psi}}_{\boldsymbol{n}}\hat{\pmb{\Psi}}_{\boldsymbol{n}}^{+}}{1+\nu}\Bigr]. 
\end{equation}
The substitution of~\eqref{inversion} into~\eqref{nu:genmus} results in
the following equation on $\nu$:
\begin{equation*}
\dfrac{(1+\nu)^2-(1+2\nu)P_0}{(1+\nu-P_0)^2}=
\dfrac1{(1-\varepsilon/2)^2},\quad P_0=\mathrm{Tr}[\mathbf{D}_0^{+}\hat{%
\boldsymbol{\Pi}}_{\boldsymbol{n}}\mathbf{D}_0],
\end{equation*}
from which follows
\begin{equation*}
\nu=P_0-1+\dfrac{(1-\varepsilon/2)\sqrt{P_0(1-P_0)}}%
{\sqrt{\varepsilon-\varepsilon^2/4}}.
\end{equation*}
The knowledge of  $\nu$ allows one  to obtain the optimum robust matrix
$\hat{\mathbf{D}}(\pmb{\theta})$ according to~\eqref{D:rob}
and, as a consequence, estimate the source position of interest
\begin{equation}  \label{MUSIC:robust}
\hat{\pmb{\theta}}= \arg\max\limits_{\pmb{\theta}}\bigl\{\mathrm{Tr}\bigl[%
\hat{\mathbf{D}}^{+}(\pmb{\theta}) \hat{\boldsymbol{\Pi}}_{\boldsymbol{n}}%
\hat{\mathbf{D}}(\pmb{\theta})]\bigr\}^{-1}.
\end{equation}

\section{Experimental results}

The experimental data was collected in the Barents Sea in October 1990 and contained receptions from a fixed sound source
located at the depth of about 148~m and emitted a narrowband signal with center frequency 240~Hz.
The propagated signal was received by a vertical array
consisting of 14 elements (equally spaced 8.5 meters apart from 44.5 m to 155 m in depth) located at the distance of $13.82$~km from the source.\footnote{%
The hydrophones located at the depths of 53, 95.5 and 155 m were not functional and were excluded in signal processing.}
The wave  roughness (wind speed) during the experiment was about $8\pm 2$~m/s.

Figure 1 models the shallow-water region of interest.
This area is characterized 
by a water depth of $\sim 170$~m above a bottom covered with silty-clay sediments.
Subsequent simulations assume that the seabed has sound speed $1780$~m/s,
density $1.8~\mbox {g/cm}^{3}$, and attenuation $0.1$~dB/wavelength.
Most of the details of that experiment may be found in~\cite{Saz02}.
\begin{figure}[ht]
\centering
\includegraphics[width=0.48\textwidth]{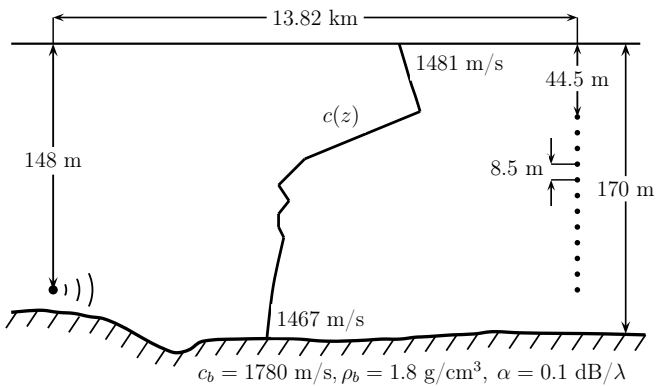}
\caption{Experimental geometry and waveguide parameters used for simulations of the localization process.}
\label{exp_geom}
\end{figure}

The received time series were quadrature demodulated, 
filtered with a bandwidth of $\pm 0.5$~Hz and subsequently spatially cross-correlated, resulting in a complex
covariance matrix. The corresponding
data covariance matrix was computed at 15 min interval from the available record using $L=1024$ snapshots.

\begin{figure}[ht]
\centering\includegraphics[width=8cm]{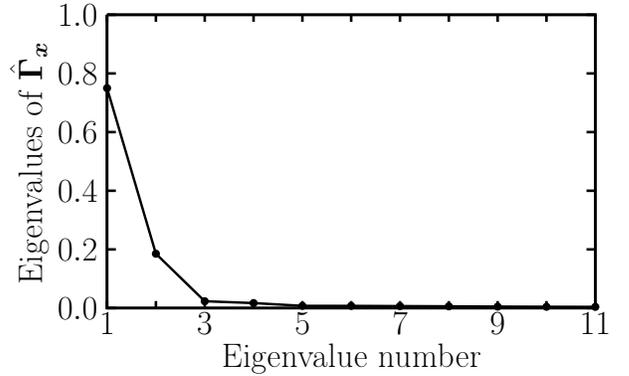}
\caption{Normalized eigenvalues of the sample correlation matrix.} 
\label{ev:240}
\end{figure}

For the experimental scenario considered, Fig.~\ref{ev:240} shows the normalized eigenvalues of $\hat{\boldsymbol{\Gamma}}_{\boldsymbol{x}}$ versus its number.
It is clearly seen from this figure that
the first two eigenvalues are predominant.
The corresponding number can be treated
as the effective dimension of signal subspace (i.e. $J=2$).
In this connection, it should be noted that
the appearance of several largest eigenvalues (prevailing over remaining)
can be explained by taking into account random sound scattering effects.

\begin{figure}[ht]
\centering\includegraphics[width=8cm]{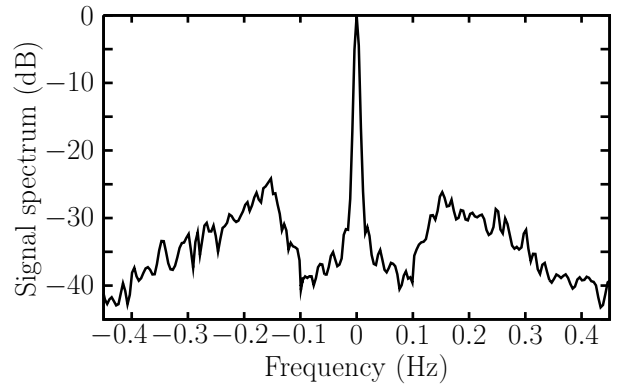}
\caption{Typical frequency spectrum of received signal in dB}
\label{spectrum}
\end{figure}

For better understanding of
the dominant mechanism of scattering in Fig.~\ref{spectrum} we plot the 
frequency spectrum (in decibel notation)
of the fluctuations observed at a fixed  array sensor depth of $138$~m.
As is seen from Fig.~\ref{spectrum} the frequency spectrum consists
of a central peak (at the carrier frequency taken as the origin)
and two lateral wings 
corresponding to the scattering component.
Such shape of the frequency spectrum is typical for a shallow water
environment where rough scattering effects are important.

The robust procedure used to estimate the source position in a random waveguide requires knowledge of
the presumed model of the signal correlation matrix ${\mathbf{R}}_{\mathbf{s}}$.
The explicit expression for the corresponding matrix in a shallow channel
in the presence of random surface scattering
is given in the Appendix.

Figure \ref{power}a illustrates the behavior of the normalized power (relative to its maximum value)
at the output of the generalized MUSIC processor utilizing the 
conventional algorithm~\eqref{MUSIC:general}.
For comparison the corresponding result for the robust case~\eqref{MUSIC:robust} 
(at $\varepsilon=0.1$)
is shown in Fig.~\ref{power}b.
\begin{figure*}[ht]
\centering
\begin{tabular}{c}
\includegraphics[width=13cm]{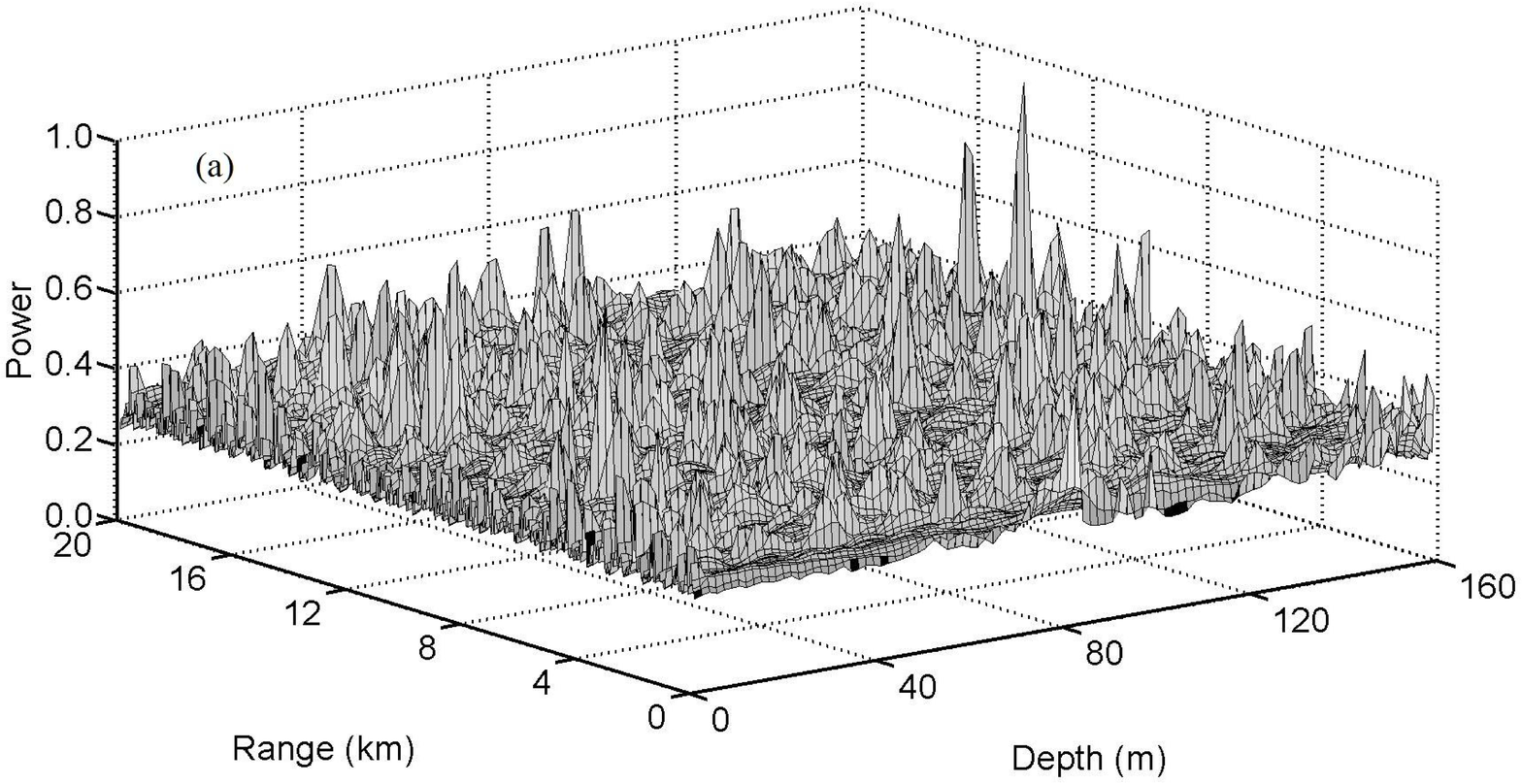} \\ 
\includegraphics[width=13cm]{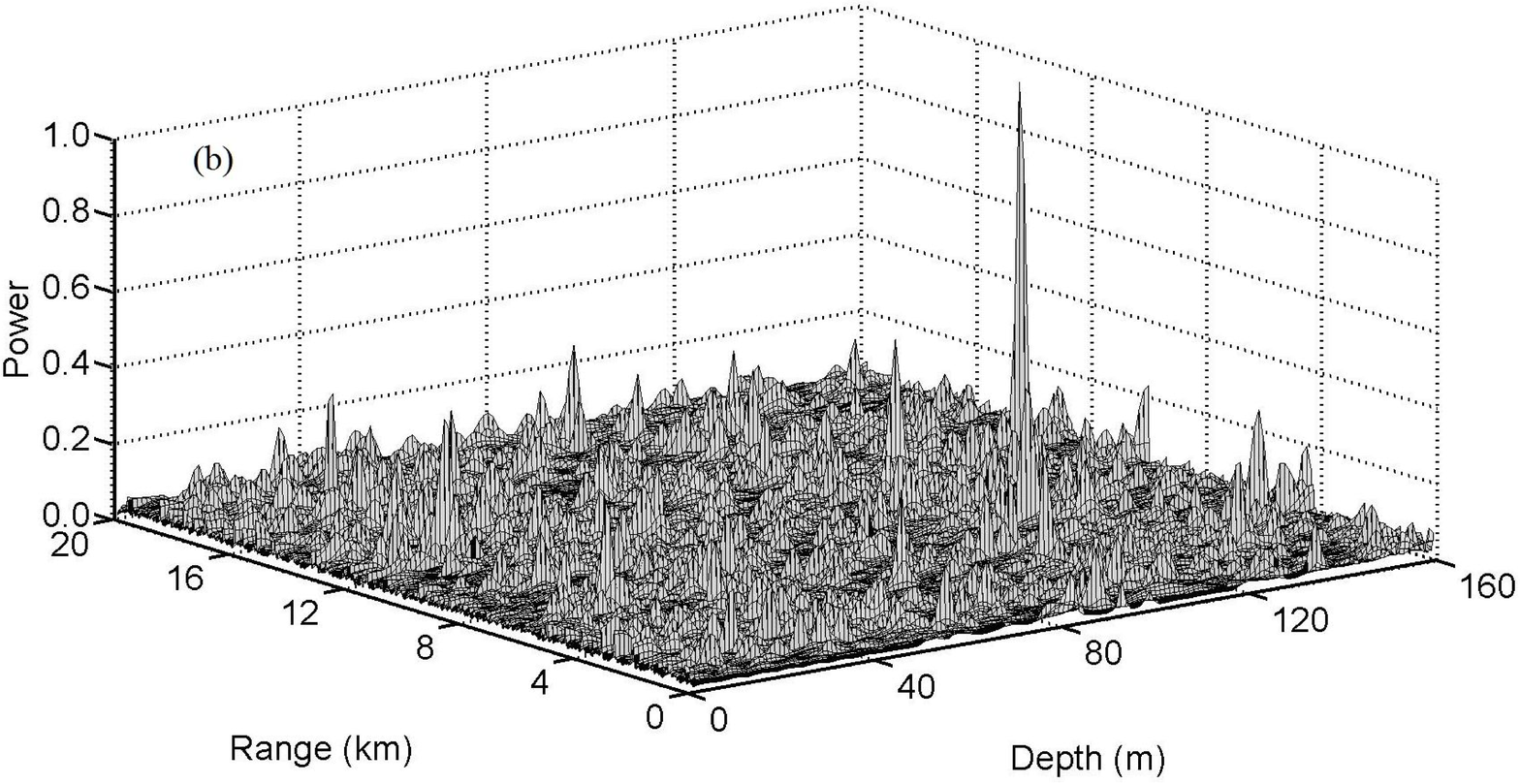}
\end{tabular}
\caption{Normalized power at the output of 
conventional (a) and robust (b) generalized MUSIC processor}
\label{power}
\end{figure*}
The output powers were computed for the hypothesized source ranges between $0$ and $20$~km at $25$ meter increments and source depths between $1$~m and $160$~m at $1$ m increments.
The wind speed used in numerical computation of the expected signal correlation matrix was taken to be 7 m/s.

One observes from Fig.~\ref{power} that both these methods produce the largest peak at  $\hat{r}_0=12.45$~km and $\hat{z}_0=147$~m that is in the neighborhood of the true source position.
However, implementation of a non-adaptive approach leads to the appearance of 
rather intensity false peaks which results in a decrease of the localization performance.

\section{Conclusion}

In this paper, we have derived a closed form robust
algorithm for estimating source location in a random shallow water channel
from the observed sample covariance matrix.
Our approach is based on the noise subspace-fitting approach
and takes into account the double constraints on the mismatched signal covariance matrix.

The algorithm is validated by its application to the experimental data observed in the Barents Sea.
For the given experimental scenario, the expected signal matrix used  in the parameter estimation problem
was predicted from the wind seas model.
The results of the real data processing demonstrate essential performance improvements that can be achieved  by means of the proposed approach.
Moreover, the presented method successfully localized the source situated at the distance of $\sim 14$~km from the array
without the need for a computationally intensive joint estimation of both the source and environmental parameters.

This work was partially supported by the Russian Foundation for Basic Research under Grants
\# 13--02--00932  and \# 13--02--97082.

\appendix
\section*{Signal correlation matrix in shallow water with  rough surface}

Consider a model of shallow water in the form of a water layer of
depth $H$ with density $\rho_w$ and sound-speed dependence $c(z)$ overlying a
semi-infinite liquid bottom of density $\rho_b$ and sound speed $c_b(1-i\alpha)$, where
$\alpha$ is a measure of the attenuation in sediment. 

To construct the presumed signal matrix ${\mathbf{R}}_{\mathbf{s}}$
used in the parameter estimation problem
we further assume  the Pierson--Moskowitz spectrum~\cite{PM}
for the rough surface spectral distribution $F(\pmb{\mbox{\ae}})$
\begin{equation}
\label{PM}
F(\pmb{\mbox{\ae}})=\frac{8.1\cdot 10^{-3}}{4\pi}\, \mbox{\ae}^{-4}\,\exp%
\Bigl(-0.74\frac{g^2}{\mbox{\ae}^2v^4}\Bigr),
\end{equation}
where  $g$ is the acceleration due to gravity, and $v$ is the wind speed over the sea surface.

In general, the correlation matrix
${\mathbf{R}}_{\mathbf{s}}$ can be expressed as
\begin{equation}  \label{R_s:def}
{\mathbf{R}}_{\mathbf{s}}(r_0,z_0)= \bigl<\mathbf{e}(r_0,z_0)\bigr>\bigl<%
\mathbf{e}^{+}(r_0,z_0)\bigr> +\mathbf{C}_{\mathbf{s}}(r_0,z_0).
\end{equation}
Here, $\bigl<\mathbf{e}(r_0,z_0)\bigr>$ is the coherent component of the signal vector
(depending on the source position of interest as on parameters)
and $\mathbf{C}_{\mathbf{s}}(r_0,z_0)$ is its covariance matrix.

In what follows we formulate the basic  formulae
for $\bigl<\mathbf{e}(r_0,z_0)\bigr>$ and $\mathbf{C}_{\mathbf{s}}(r_0,z_0)$
in a shallow water channel (where the combined effects of rough surface scattering and bottom interactions are important)
based on the results obtained in Ref.~\cite{Saz02}.

In the framework of a modal approach, the vector components
$\left\{\bigl<e_j(r_0,z_0)\bigr>\right\}_{j=1}^{N}$ 
can be represented as a sum over a total number $M$ of the propagating modes
\begin{equation}  \label{coher}
\bigl<e_j(r_0,z_0)\bigr> =\overline{I}_s^{\,-1/2}\sum\limits_{n=1}^{M}
\dfrac{\varphi_n(z_0)\varphi_n(z_j)}{\sqrt{\kappa_m}}\, e^{\textstyle\,
(i\kappa_n-0.5\sigma_n)r_0}, \\ 
\end{equation}
Here, $\varphi_{n}(z)$ and $\kappa_n$ are,  respectively, the depth eigenfunction and the horizontal wavenumber of the $n$-th mode,
$\sigma_n$ is the modal attenuation parameter describing the combined
effects of absorption and scattering losses:
$\sigma_n=\sigma_n^a+\sigma_n^s$,
and $\overline{I}_s$ is the normalization factor determined from the condition
$\Tr{\bf R}_{\mathbf{s}}=1$.

In the modal representation the covariance  matrix elements
$[\mathbf{C}_{\mathbf{s}}(r_0,z_0)]_{jk}$ 
are given by the expression
\begin{multline}
\label{C_s:cov}
[\mathbf{C}_{\mathbf{s}}(r_0,z_0)]_{jk}=\dfrac1{\overline{I}_s}
\sum\limits_{n=1}^{M}\dfrac1{\kappa_n} \left[\,I_{n}(r_0,z_0)\phantom{e^{\textstyle -\sigma_n r_0}}\right.\\
-\left.
\varphi_n^2(z_0)\,e^{\textstyle -\sigma_n r_0}\right]\varphi_n(z_j)\varphi_n(z_k),
\end{multline}
where the quantity $I_{n}(r_0,z_0)$ (having the sense of modal intensity) 
obeys the transport equation
\begin{equation}  \label{master}
\left(\frac{d}{dr_0}+\sigma_n^a\right)I_{n}(r_0,z_0)= \sum\limits_{m=1}^M
a_{nm}\Bigl[I_{m}(r_0,z_0)-I_{n}(r_0,z_0)\Bigr]
\end{equation}
governing the change of the  modal intensity as a result of random scattering and bottom absorption.

For the Pierson--Moskowitz distribution, Eq.~(\ref{PM}), the calculation
of the coupling matrix $a_{nm}$ is given in~\cite{Beil}. The result is:
\begin{gather*}
a_{nm}=\frac{8.1\!\cdot\!10^{-3}\sqrt{2}\pi
\left[\varphi_n^{\prime}(0)\varphi_m^{\prime }(0)\right]^2}{8\,\kappa_n\kappa_m k_0^3}\,
f(x_{nm}),\\
 f(x)=x^{3/2}e^{-x}\left[I_0(x)-I_1(x)\right],\ x_{nm}=\dfrac{%
0.5\,k_0^2}{\left(\kappa_n\!-\!\kappa_m\right)^2}.
\end{gather*}
Here, $k_0^2=0.74g^2/v^4$,
$x_{nm}=0.5\,k_0^2/\left(\kappa_n\!-\!\kappa_m\right)^2$,
and
$$
f(x)=x^{3/2}e^{-x}\left[I_0(x)-I_1(x)\right],
$$
where  $I_0$ and $I_1$ are the modified Bessel functions of order zero and
unity, respectively, 
and the prime  denotes  differentiation  with respect  to depth $z$.

The solution of Eq.~(\ref{master}) has the form
\begin{equation}  \label{rte:sol}
\begin{split}
I_n(r_0,z_0)=&\sum\limits_{m=1}^{M} g_{nm}(r_0)\varphi_m^2(z_0),\\
g_{nm}(r_0)=&\sum_{l=1}^{M}\Phi_n^{(l)}\, e^{\textstyle -\lambda_l\,r_0}\,%
\Phi_m^{(l)},
\end{split}
\end{equation}
where the functions $\Psi_{n}^{(l)}$ are the eigenvectors of the matrix $\lVert\delta_{nm}\,\sigma_n-a_{nm}\rVert$,
associated with the eigenvalues  $\lambda_l$.

Notice that the scattering coefficient $\sigma_n^s$  is  related to the coupling matrix $a_{nm}$
by $\sigma_n^s=\sum\limits_{m=1}^Ma_{nm}$.
As for the modal attenuation parameter $\sigma_n^{\,a}$, for a liquid bottom
the corresponding coefficient can be obtained from the perturbation theory
(see, e.g.~\cite{Kats}):
\begin{equation*}
\sigma_n^a=\frac{\rho_w}{\rho_b}\,
\frac{k^2n_{b}^2|\varphi_n(H)|^2}{\kappa_n \sqrt{\kappa_n^2-k^2n_{b}^2}}\,\alpha,
\end{equation*}
where $n_b=c(H)/c_b$ is the bottom refraction index
and $k$ is the reference wavenumber.

For a given set of parameters (sound profile, wind speed and source frequency)
Eqs.~(\ref{R_s:def})--(\ref{C_s:cov}) together with~(\ref{rte:sol})
give an explicit rule for calculating the signal correlation matrix 
versus source position.

\end{document}